\documentclass[aps,prl,reprint,groupedaddress]{revtex4-2}

\usepackage{graphicx}
\usepackage{dcolumn}
\usepackage{bm}
\usepackage{amsmath}
\usepackage{subfigure}

\begin{document}


\title{Passive scalar dispersion along porous stratum with natural convection}


\author{Chenglong Hu}
\affiliation{State Key Laboratory for Turbulence and Complex Systems, and Department of Mechanics and Engineering Science, College of Engineering, Peking University, Beijing 100871, P.R. China}

\author{Ke Xu}
\email{kexu1989@pku.edu.cn}
\affiliation{Department of Energy and Resources Engineering, College of Engineering, Peking University, Beijing 100871, P.R. China}
\affiliation{Institute of Energy, Peking University, Beijing 100871, P.R. China}

\author{Yantao Yang}
\email[]{yantao.yang@pku.edu.cn}
\affiliation{State Key Laboratory for Turbulence and Complex Systems, and Department of Mechanics and Engineering Science, College of Engineering, Peking University, Beijing 100871, P.R. China}
\affiliation{Laoshan Laboratory, Qingdao 266299, Shandong, P.R. China}


\date{\today}

\begin{abstract}
  We investigate horizontal dispersion of a passive scalar in a porous stratum with Rayleigh-Darcy convection initiated by a geothermal gradient. While increasing Rayleigh number ($Ra$) keeps enhancing convection, the horizontal dispersion coefficient ($\hat{D}$) only scales with Ra in a narrow range, and saturates at $Ra \gtrsim 2500$. We rationalize this two-stage behavior: at low Ra, passive scalar migrates through bulk circumflex; at high Ra, boundary micro-plume layer becomes the dominant scalar pathway. Theory gives saturate $\hat{D}$ value proportional to the Lewis number $Le$ and molecular diffusivity $D_0$.
\end{abstract}


\maketitle


Passive scalars (solute, colloids suspensions, etc.) migrate along subsurface porous strata, which brings long-range environmental and geochemical impacts \cite{HUNT_Review, KANTISEN200671}. Examples include radioactive contaminant spreading \cite{radioactive_science, 10.1007/978-0-387-88483-7_24}, solute exchange between neighboring surface aquifers \cite{WANG201351}, leakage of petroleum chemicals\cite{TARTAKOVSKY2013247}, geochemical tracer tests \cite{Moreno1985951}, and complicated long-range subsurface microbe and colloid migration\cite{Tufenkji2007, MicrobeNPAnna, colloid_review}. Investigation of these key processes requires the correct description of horizontal solvent dispersion dynamics along a porous medium.

However, horizontal solute transport is not trivial in a subsurface stratum. Intrinsic geothermal gradient may largely complicate it, even in the absence of additional non-idealities such as heterogeneity, inclinations and background flow. Fluid near the lower boundary is warmer and thus of lower density, while fluid near the upper boundary is cooler and thus of higher density. The Rayleigh number, defined as the ratio between gravitational flow flux and molecular diffusion flux, can vary up to $\mathcal{O}(10^4)$ for geothermal systems~\cite{hewitt2020vigorous} and $\mathcal{O}(10^5 \sim 10^6)$ for geological carbon dioxide sequestration~\cite{pirozzoli2021ultimate,depaoli2022strong}. Natural convection in porous media, denoted as Rayleigh-Darcy convection, can thus be initiated \cite{PhysRevLett.109.264503} and evolve to a statistically steady-state \cite{hewitt2012ultimate}. 

Natural convection largely enhances vertical mass and heat transport. In two-dimensional (2D) simulations, the vertical transport rate (characterized by the Nusselt number $Nu$) scales with $Ra$ as $Nu \sim Ra$ when $500 \lesssim  Ra \lesssim 1200$, and as $Nu \sim Ra^{0.9}$ when $ 1255 \le Ra \le 10^4$~\cite{otero2004highrayleighnumber}. A linear scaling is attained asymptotically with increasing $Ra$ up to $4\times10^4$~\cite{hewitt2012ultimate}. The literature also emphasizes how boundary conditions, heterogeneity, fluid properties and reactive behaviors shapes convection ~\cite{slim2014solutalconvection,salibindla2018dissolutiondriven,daniel2014effect,hidalgo2015dissolution,fu2015rock}. Unfortunately, regardless of extensive studies on vertical mass and heat transport during Rayleigh-Darcy convection, very few studies look into the horizontal dispersion dynamics.   

Here we investigate how natural convection driven by vertical gradients regulates horizontal dispersion of a passive scalar (also referred to as concentration or solute) in a porous stratum. We consider 2D Rayleigh-Darcy convection within a fluid-saturated porous medium with uniform porosity $\phi$ and permeability $K$. Denote the height and the width of flow domain by $H$ and $L$. We set $H \ll L$ to match practical condition and to obtain  statistical convergence in the horizontal solute transport. Let $(x,z)$ be the horizontal and vertical coordinates, and $\mathbf{u}=(u,w)$ the velocity and corresponding components, respectively. Both the temperature $T$ and concentration $S$ of the incompressible Darcy flow follow the standard advection-diffusion equation~\cite{hewitt2020vigorous}. The density of fluid decreases linearly with temperature, resulting in an unfavorable vertical density contrast $\Delta\rho_T$ that drives the convection within the layer. At the top and bottom boundaries, the temperature is constant and for concentration field the adiabatic or no-flux condition is applied. By using $U=Kg \Delta \rho_T/\mu$ as the characteristic scale for velocity, $H$ for length, and $t_c=\phi H/U$ for time, the control parameters include the Rayleigh number $Ra=KHg \Delta \rho_T /\phi\mu\kappa_T$ measuring the strength of the convection motions, and the Lewis number $Le=\kappa_T/\kappa_S$. Here $\mu$ is the dynamic viscosity of the fluid, $\kappa_T$ is the thermal diffusivity of the fluid, and $\kappa_S$ is the molecular diffusivity of the passive scalar. The complete theoretical formulation is provided in the Supplementary Material~\footnote{See Supplemental Material at [URL will be inserted by publisher] for numerical methods.} and the numerical method is the same as that in our previous work~\cite{hu2023effects}.

A series of simulations are carried out with $Ra$ ranging from $10$ up to $2\times10^4$ for $Le=2$. Once the convection flow reaches the statistically steady state, the vertically aligned column of concentration field is introduced into the flow field with an initial Gaussian distribution of $S_0(x)=\exp\left[-(x-x_0)^2/(4\tau)\right]$. Here $x_0$ denotes the center location of the initial distribution and $\tau$ is set so that $S_0(x_0\pm 0.15L/H)=0.01$. To quantify the horizontal solute dispersion coefficient $\hat{D}$, we assume that vertically averaged concentration $\tilde{S}$ follows the 1D diffusion equation $\partial_t \tilde{S} = \hat{D} \partial^2\tilde{S}/\partial x^2$. The above diffusion equation has the analytic solution
\begin{equation}\label{eqn:extS}
  \tilde{S}(x,t)=\sqrt{\frac{\tau}{\hat{D}t+\tau}}\exp{\left(\frac{-(x-x_0)^2}{4(\hat{D}t+\tau)}\right)}
\end{equation}
for an initial distribution $S_0(x)$. At any given time,  $\hat{D}$ can be numerically determined by fitting the vertically averaged concentration field against the analytic solution~\eqref{eqn:extS}. For each $Ra$, we run five cases with equidistant $x_0$ separated by $0.2 L/H$ to reduce uncertainty. 

\begin{figure}
  \centering
  \includegraphics[width=0.49\textwidth]{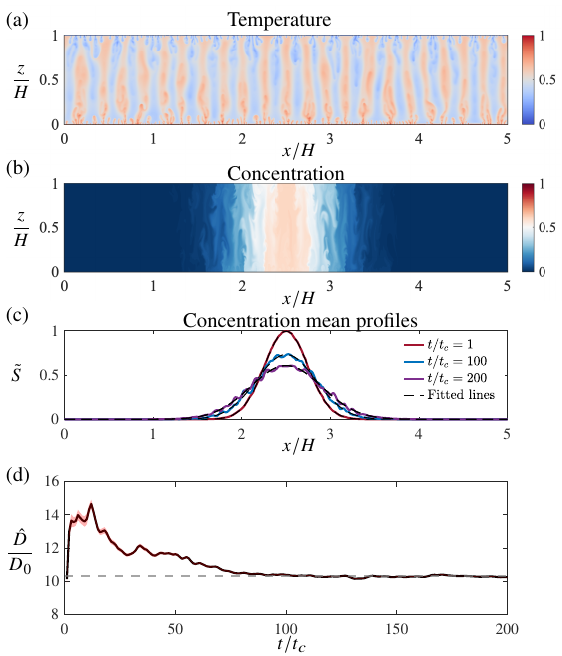}
  \caption{\label{fig:RaT2e4}Snapshots of (a) temperature and (b) concentration fields $t/t_c=200$ with an initial center location $x_0=0.5L/H$ for $Ra=2\times10^4,~Le=2$. (c) Concentration mean profiles (solid lines) at different time with their fitted lines (dashed lines) by the equation \eqref{eqn:extS}. (d) Variations of the fitted $\hat{D}/D_0$ with time. The gray dashed line denotes the value averaged over the last 100 time period. The bounded red shadow denotes the $95\%$ confidence interval for the fitted coefficient.}
\end{figure}

Fig.~\ref{fig:RaT2e4}(a) demonstrates the temperature field at $t/t_c=200$ for case $Ra=2\times10^4$, where the turbulent convection flow motions can be identified. Fig.~\ref{fig:RaT2e4}(b) demonstrates the concentration field at the same time, and the turbulent diffusion in the horizontal direction can be clearly seen. The horizontal profiles $\tilde{S}(x,t)$ given by the vertical average are plotted in Fig.~\ref{fig:RaT2e4}(c) at several different times. All the curves exhibit a Gaussian shape, which validates the use of 1D diffusion equation. The temporal variation of rescaled fitted dispersion coefficient  $\hat{D}/D_0$ is presented in Fig.~\ref{fig:RaT2e4}(d) where $D_0=\phi \kappa_S$ is the molecular diffusivity of concentration in porous media. After the initial transition stage, $\hat{D}/D_0$ keeps nearly constant until the end of simulation. The dispersion coefficient is then determined for the given $Ra$ by first averaging over the final statistically steady stage and then over the five runs with different initial location $x_0$. 

\begin{figure}
  \centering
  \includegraphics[width=0.45\textwidth]{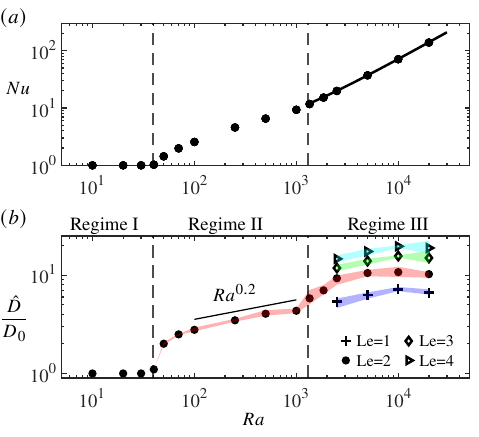}
  \caption{\label{fig:Dfit} (a) Variations of the Nusselt number with $Ra$. The solid line denotes a linear relation $Nu=0.0069Ra+2.75$ for $Ra\ge 1300$ given by~\cite{hewitt2012ultimate}. (b) Variations of the rescaled dispersion coefficient $\hat{D}/D_0$ with $Ra$, where the symbols and shadows are mean values and value ranges of coefficients at five different $x_0$, respectively. The left dashed line denotes the critical Rayleigh number $Ra_c=4\pi^2$, above which convective motions happen. The right dashed line marks the transition to high-$Ra$ regime.}
\end{figure}

 As shown in Fig.~\ref{fig:Dfit}, we characterize the evolution of vertical and horizontal transport performance under varying $Ra$. Fig.~\ref{fig:Dfit}(a) shows the evolution of the thermal flux $Nu$, defined as the averaged dimensionless vertical temperature gradient at the top and bottom boundaries~\cite{hewitt2012ultimate}. The thermal flux keeps increasing when $Ra$ exceeds the critical value $Ra\ge 4\pi^2$. In contrast, the horizontal dispersion coefficient $\hat{D}/D_0$ in Fig.~\ref{fig:Dfit}(b), does not always increase with $Ra$. An astonishing observation is that $\hat{D}/D_0$ reaches a saturate value for $Ra\gtrsim 2500$.

The evolution of $\hat{D}$ can be divided into three regimes. In regime \uppercase\expandafter{\romannumeral1} where $Ra<4\pi^2$, there is no convection motion, and $\hat{D}/D_0$ is close to unit corresponding to molecular diffusion. This is referred to as the diffusion regime. In regime \uppercase\expandafter{\romannumeral2} that $Ra$ becomes larger than $4\pi^2$, $\hat{D}/D_0$ rapidly jumps up as the convection motions develop, and then increases with $Ra$ by $\hat{D}\sim Ra^{0.2}$. As $Ra$ further increases over $1300$ in regime \uppercase\expandafter{\romannumeral3}, there appears a second transition after which $\hat{D}/D_0$ stabilizes at an ultimate saturate value close to $10$ for $Le=2$, that no longer increases with $Ra$.

\begin{figure}
  \centering
  \includegraphics[width=0.4\textwidth]{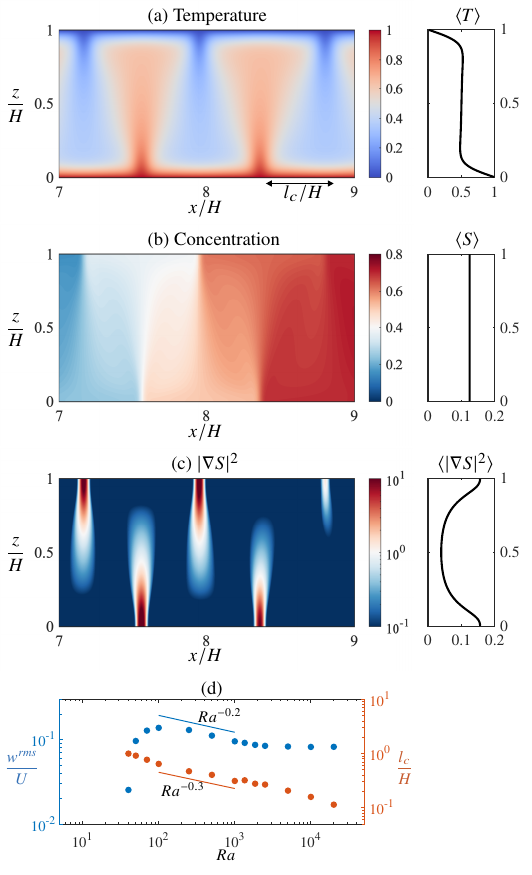}
  \caption{\label{fig:cell} The transient snapshots and corresponding horizontally averaged mean profiles of (a) temperature, (b) concentration and (c) concentration dissipation for a typical Rayleigh number $Ra=500,~Le=2$ in regime \uppercase\expandafter{\romannumeral2} at $t/t_c=100$, respectively. (d) Variations of the r.m.s. vertical velocity at middle height (left axis) and the characteristic length of convection cells (right axis) obtained from autocorrelation function. The arrowed line in (a) represents the cell length $l_c$.}
\end{figure}

The transition $Ra$ for three horizontal dispersion regimes well corresponds with those of vertical flux evolution. It implies that horizontal dispersion pattern is closely correlated to the convection pattern. We model the dimensional horizontal diffusivity as $\hat{D} \sim \mathcal{U}\mathcal{L}$ with $\mathcal{U}$ and $\mathcal{L}$ being the appropriately chosen characteristic scales for horizontal velocity and length, respectively. To rationalize the observed $\hat{D}$ evolution with $Ra$, we need to determine the corresponding horizontal velocity and length scale in different regimes. 
 
 For regime \uppercase\expandafter{\romannumeral2} that $10^2<Ra<10^3$, the flow field consists of large-scale quasi-periodic convection rolls with nearly no spatial variations over time. Fig.~\ref{fig:cell}(a) illustrates a few roll structures for $Ra=500$ with scalar concentration in (b). The concentration dissipation in (c) measured by $|\nabla S|^2$ attains it maximum on the boundaries and extends vertically to the interior. A clear circulation of ascending hot fluid and descending cold fluid contributes to the dispersion of concentration, also see the supplemental movie~\footnote{See Supplemental Material at [URL will be inserted by publisher] for a movie of temperature and concentration at $Ra=500$.}. Therefore, we refer this regime to the cell-dominated regime. By the flux conservation within one convection cell, the transverse velocity should scale as $\mathcal{U}\sim\mathcal{W} l_c / H$. Here $\mathcal{W}$ is the characteristic scale of vertical velocity and $l_c$ is the width of cell~\footnote{See Supplemental Material at [URL will be inserted by publisher] for calculating the width of cell $l_c$.}. For convection cells the vertical velocity scale can be readily chosen as $\mathcal{W}\sim w^\mathit{rms}$ and is directly related to the driving buoyancy force. Indeed, as shown in Fig.~\ref{fig:cell}(d), a power-law scaling $w^\mathit{rms}\sim Ra^{-0.2} U \sim Ra^{0.8}$ clearly exists in the cell-dominated regime with roughly $10^2<Ra<10^3$. Furthermore, the cell width scales as $l_c /H \sim Ra^{-0.3}$. By combining the above scaling relations, one obtains 
\begin{equation}
    \hat{D} \sim \mathcal{U}\mathcal{L} 
    \sim \mathcal{W} \frac{l_c}{H} l_c 
    \sim Ra^{0.2}.
\end{equation}
This scaling relation for the cell-dominated regime is plotted in Fig.~\ref{fig:Dfit}(b) and in excellent agreement with the numerical experiments.

The most intriguing observation in this study is that the horizontal diffusivity saturates in regime \uppercase\expandafter{\romannumeral3} at large $Ra$, regardless of continuous amplification of vertical convection. The transition $Ra$ between regime \uppercase\expandafter{\romannumeral2} and regime \uppercase\expandafter{\romannumeral3} corresponds to the transition in the flow morphology from the nearly steady large convection rolls to an unsteady multi-scale state. This multi-scale state is similar to the previous work on the 2D Rayleigh-Darcy convection in the ``high-$Ra$'' regime~\cite{hewitt2012ultimate}. Namely, small ``protoplumes'' originate from boundaries and drive small local circulations between them near the boundary. These protoplumes merge into vigorous columnar ``megaplumes'' which in term form the large scale rolls in the bulk. This multi-scale state is illustrated in Fig.~\ref{fig:plume}(a-c), see also the supplementary movies~\footnote{See Supplemental Material at [URL will be inserted by publisher] for a movie of temperature and concentration at $Ra=2\times10^4$.}.

\begin{figure}
  \centering
  \includegraphics[width=0.48\textwidth]{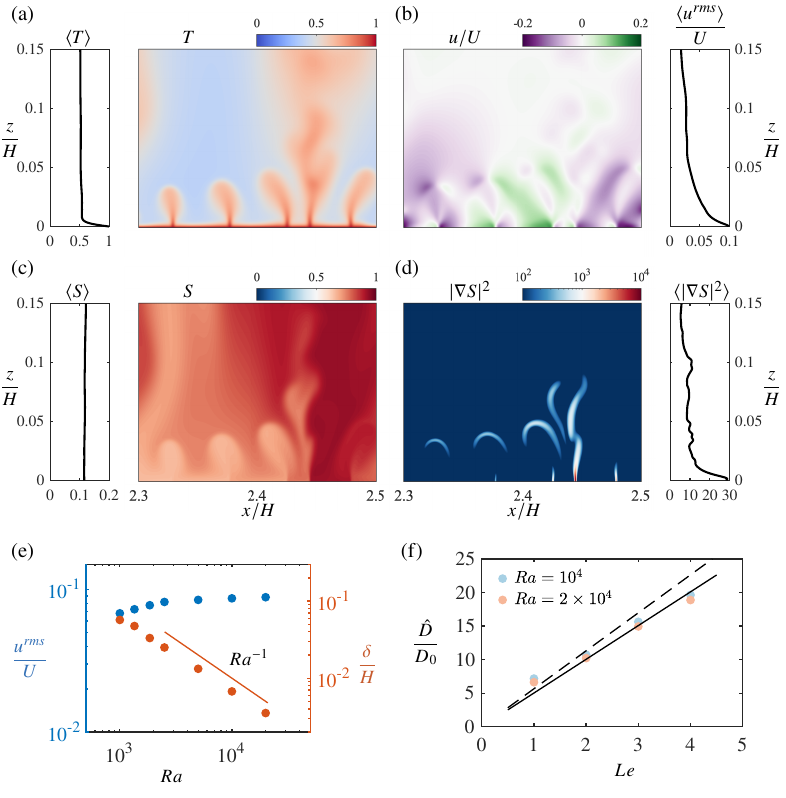}
  \caption{\label{fig:plume} The close-up transient snapshots and corresponding horizontally averaged mean profiles of (a) temperature, (b) horizontal velocity, (c) concentration and (d) concentration dissipation for a typical Rayleigh number $Ra=2\times 10^4,~Le=2$ in regime \uppercase\expandafter{\romannumeral3} at $t/t_c=20$, respectively. (e) Variations of the r.m.s. horizontal velocity at boundaries (left axis) and the characteristic length of protoplumes (right axis) with $Ra$. (f) Variations of the rescaled dispersion coefficient with the Lewis number at different Rayleigh numbers. The dashed line denotes the modeled linear relation \eqref{eqn:regime3} and the solid line is a linear fit with a slope of 5.}
\end{figure}

In Fig.~\ref{fig:plume}(b) and Fig.~\ref{fig:plume}(d), we observe that major horizontal advection and horizontal concentration dissipation are concentrated in the near-boundary micro-plume region. Therefore, we conclude that the horizontal mixing of the passive scalar is controlled by the small circulations between protoplumes near boundary, instead of the large scale rolls driven by the megaplumes. We refer to this high-$Ra$ regime as the plume-dominated regime.

A natural choice of the horizontal length scale is the boundary layer thickness $\delta$. Established theory shows that $\delta \sim Ra^{-1}$, with a pre-factor 50 in 3D scenario~\cite{zhu2024transporta}. In our 2D simulation, $\delta$ is determined as 66.7 by measuring the location of the intersection between two linearly fitted lines of temperature profile in the interior and near-wall regions~\footnote{See Supplemental Material at [URL will be inserted by publisher] for the illustration of determining boundary layer thickness.}. The velocity scale is the averaged horizontal r.m.s. velocity $u^\mathit{rms}$ within  boundary layers, which is near a constant value divided by $U$. Our numerical 
results (Fig~\ref{fig:plume}(e)) show that $u^\mathit{rms}/U\approx0.085$. It can be simply rationalized as that the fluid velocity in the boundary layer is consistent with that in the bulk region, which has already been reported as 0.082~\cite{hewitt2012ultimate}. 

The dispersion coefficient is thus modeled as
\begin{equation}\label{eqn:regime3}
  \hat{D} = u^\mathit{rms} \delta \approx 0.085 Ra Le D_0 \frac{\delta}{H} =  5.66 Le D_0.
\end{equation}
The pre-factor of 5.66 results from the combined effects of the horizontal velocity magnitude and thermal boundary layer thickness. We validate such a relation by varying the Lewis number $Le$ from 1 to 4 for $Ra\ge 10^4$. The results in Fig.~\ref{fig:plume}(f) are in good agreements with the presumed relation \eqref{eqn:regime3}, and a pre-factor of 5 provides a more accurate description.

In summary, we investigate horizontal dispersion of a passive scalar under the Rayleigh-Darcy convection in a porous medium. We identify how convection pattern determines the horizontal dispersion mode. For the intermediate Rayleigh number, the convection is dominated by large scale rolls which extend over the entire domain height, so horizontal dispersion is controlled by the characteristic horizontal velocity and width of these rolls, which results in a $\hat{D} \sim Ra^{0.2}$ scaling. For higher Rayleigh numbers, the horizontal mixing is no longer dominated by the large scale rolls, but rather by the small scale circulations between the protoplumes near the boundary. The boundary layer shrinks under increased $Ra$, which offsets the simultaneous increasing of velocity. It results in a surprising saturate $\hat{D}$ value that does not change with $Ra$, although the turbulent convection strength continuously scales with $Ra$. The saturate value is determined by Lewis number and molecular diffusivity as $\hat{D} = 5 Le D_0$ in 2D stratum.  

Future research would focus on how more complicated boundary conditions and background flow may affect the horizontal dispersion behaviors in porous stratum.

\bigskip
\begin{acknowledgments}
CH and YY acknowledge the financial support from the National Natural Science Foundation of China under the Grant 11988102 and the Laoshan Laboratory under the Grant No.LSKJ202202000. KX acknowledges the financial support from the National Key Research and Development Project of China under the Grant No.2023YFA1011700 and from the China National Petroleum Corporation-Peking University Strategic Cooperation Project of Fundamental Research. 
\end{acknowledgments}

\bibliography{transport}

\begin{thebibliography}{28}%
\makeatletter
\providecommand \@ifxundefined [1]{%
 \@ifx{#1\undefined}
}%
\providecommand \@ifnum [1]{%
 \ifnum #1\expandafter \@firstoftwo
 \else \expandafter \@secondoftwo
 \fi
}%
\providecommand \@ifx [1]{%
 \ifx #1\expandafter \@firstoftwo
 \else \expandafter \@secondoftwo
 \fi
}%
\providecommand \natexlab [1]{#1}%
\providecommand \enquote  [1]{``#1''}%
\providecommand \bibnamefont  [1]{#1}%
\providecommand \bibfnamefont [1]{#1}%
\providecommand \citenamefont [1]{#1}%
\providecommand \href@noop [0]{\@secondoftwo}%
\providecommand \href [0]{\begingroup \@sanitize@url \@href}%
\providecommand \@href[1]{\@@startlink{#1}\@@href}%
\providecommand \@@href[1]{\endgroup#1\@@endlink}%
\providecommand \@sanitize@url [0]{\catcode `\\12\catcode `\$12\catcode
  `\&12\catcode `\#12\catcode `\^12\catcode `\_12\catcode `\%12\relax}%
\providecommand \@@startlink[1]{}%
\providecommand \@@endlink[0]{}%
\providecommand \url  [0]{\begingroup\@sanitize@url \@url }%
\providecommand \@url [1]{\endgroup\@href {#1}{\urlprefix }}%
\providecommand \urlprefix  [0]{URL }%
\providecommand \Eprint [0]{\href }%
\providecommand \doibase [0]{https://doi.org/}%
\providecommand \selectlanguage [0]{\@gobble}%
\providecommand \bibinfo  [0]{\@secondoftwo}%
\providecommand \bibfield  [0]{\@secondoftwo}%
\providecommand \translation [1]{[#1]}%
\providecommand \BibitemOpen [0]{}%
\providecommand \bibitemStop [0]{}%
\providecommand \bibitemNoStop [0]{.\EOS\space}%
\providecommand \EOS [0]{\spacefactor3000\relax}%
\providecommand \BibitemShut  [1]{\csname bibitem#1\endcsname}%
\let\auto@bib@innerbib\@empty
\bibitem [{\citenamefont {Hunt}\ and\ \citenamefont
  {Sahimi}(2017)}]{HUNT_Review}%
  \BibitemOpen
  \bibfield  {author} {\bibinfo {author} {\bibfnamefont {A.~G.}\ \bibnamefont
  {Hunt}}\ and\ \bibinfo {author} {\bibfnamefont {M.}~\bibnamefont {Sahimi}},\
  }\bibfield  {title} {\bibinfo {title} {Flow, transport, and reaction in
  porous media: Percolation scaling, critical-path analysis, and effective
  medium approximation},\ }\href
  {https://doi.org/https://doi.org/10.1002/2017RG000558} {\bibfield  {journal}
  {\bibinfo  {journal} {Reviews of Geophysics}\ }\textbf {\bibinfo {volume}
  {55}},\ \bibinfo {pages} {993} (\bibinfo {year} {2017})}\BibitemShut
  {NoStop}%
\bibitem [{\citenamefont {{Kanti Sen}}\ and\ \citenamefont
  {Khilar}(2006)}]{KANTISEN200671}%
  \BibitemOpen
  \bibfield  {author} {\bibinfo {author} {\bibfnamefont {T.}~\bibnamefont
  {{Kanti Sen}}}\ and\ \bibinfo {author} {\bibfnamefont {K.~C.}\ \bibnamefont
  {Khilar}},\ }\bibfield  {title} {\bibinfo {title} {Review on subsurface
  colloids and colloid-associated contaminant transport in saturated porous
  media},\ }\href {https://doi.org/https://doi.org/10.1016/j.cis.2005.09.001}
  {\bibfield  {journal} {\bibinfo  {journal} {Advances in Colloid and Interface
  Science}\ }\textbf {\bibinfo {volume} {119}},\ \bibinfo {pages} {71}
  (\bibinfo {year} {2006})}\BibitemShut {NoStop}%
\bibitem [{\citenamefont {Winogard}(1981)}]{radioactive_science}%
  \BibitemOpen
  \bibfield  {author} {\bibinfo {author} {\bibfnamefont {I.~J.}\ \bibnamefont
  {Winogard}},\ }\bibfield  {title} {\bibinfo {title} {Radioactive waste
  disposal in thick unsaturated zones},\ }\href
  {https://doi.org/10.1126/science.212.4502.1457} {\bibfield  {journal}
  {\bibinfo  {journal} {Science}\ }\textbf {\bibinfo {volume} {212}},\ \bibinfo
  {pages} {1457} (\bibinfo {year} {1981})}\BibitemShut {NoStop}%
\bibitem [{\citenamefont {Jin}\ and\ \citenamefont
  {Chang}(2009)}]{10.1007/978-0-387-88483-7_24}%
  \BibitemOpen
  \bibfield  {author} {\bibinfo {author} {\bibfnamefont {A.}~\bibnamefont
  {Jin}}\ and\ \bibinfo {author} {\bibfnamefont {S.-Y.}\ \bibnamefont
  {Chang}},\ }\bibfield  {title} {\bibinfo {title} {Radioactive contaminanant
  transport in subsurface porous environment},\ }in\ \href@noop {} {\emph
  {\bibinfo {booktitle} {Proceedings of the 2007 National Conference on
  Environmental Science and Technology}}},\ \bibinfo {editor} {edited by\
  \bibinfo {editor} {\bibfnamefont {E.}~\bibnamefont {Nzewi}}, \bibinfo
  {editor} {\bibfnamefont {G.}~\bibnamefont {Reddy}}, \bibinfo {editor}
  {\bibfnamefont {S.}~\bibnamefont {Luster-Teasley}}, \bibinfo {editor}
  {\bibfnamefont {V.}~\bibnamefont {Kabadi}}, \bibinfo {editor} {\bibfnamefont
  {S.-Y.}\ \bibnamefont {Chang}}, \bibinfo {editor} {\bibfnamefont
  {K.}~\bibnamefont {Schimmel}},\ and\ \bibinfo {editor} {\bibfnamefont
  {G.}~\bibnamefont {Uzochukwu}}}\ (\bibinfo  {publisher} {Springer New York},\
  \bibinfo {address} {New York, NY},\ \bibinfo {year} {2009})\ pp.\ \bibinfo
  {pages} {181--188}\BibitemShut {NoStop}%
\bibitem [{\citenamefont {Wang}\ and\ \citenamefont {Zhan}(2013)}]{WANG201351}%
  \BibitemOpen
  \bibfield  {author} {\bibinfo {author} {\bibfnamefont {Q.}~\bibnamefont
  {Wang}}\ and\ \bibinfo {author} {\bibfnamefont {H.}~\bibnamefont {Zhan}},\
  }\bibfield  {title} {\bibinfo {title} {Radial reactive solute transport in an
  aquifer–aquitard system},\ }\href
  {https://doi.org/https://doi.org/10.1016/j.advwatres.2013.08.013} {\bibfield
  {journal} {\bibinfo  {journal} {Advances in Water Resources}\ }\textbf
  {\bibinfo {volume} {61}},\ \bibinfo {pages} {51} (\bibinfo {year}
  {2013})}\BibitemShut {NoStop}%
\bibitem [{\citenamefont {Tartakovsky}(2013)}]{TARTAKOVSKY2013247}%
  \BibitemOpen
  \bibfield  {author} {\bibinfo {author} {\bibfnamefont {D.~M.}\ \bibnamefont
  {Tartakovsky}},\ }\bibfield  {title} {\bibinfo {title} {Assessment and
  management of risk in subsurface hydrology: A review and perspective},\
  }\href {https://doi.org/https://doi.org/10.1016/j.advwatres.2012.04.007}
  {\bibfield  {journal} {\bibinfo  {journal} {Advances in Water Resources}\
  }\textbf {\bibinfo {volume} {51}},\ \bibinfo {pages} {247} (\bibinfo {year}
  {2013})},\ \bibinfo {note} {35th Year Anniversary Issue}\BibitemShut
  {NoStop}%
\bibitem [{\citenamefont {Moreno}\ \emph {et~al.}(1985)\citenamefont {Moreno},
  \citenamefont {Neretnieks},\ and\ \citenamefont {Eriksen}}]{Moreno1985951}%
  \BibitemOpen
  \bibfield  {author} {\bibinfo {author} {\bibfnamefont {L.}~\bibnamefont
  {Moreno}}, \bibinfo {author} {\bibfnamefont {I.}~\bibnamefont {Neretnieks}},\
  and\ \bibinfo {author} {\bibfnamefont {T.}~\bibnamefont {Eriksen}},\
  }\bibfield  {title} {\bibinfo {title} {Analysis of some laboratory tracer
  runs in natural fissures},\ }\href {https://doi.org/10.1029/WR021i007p00951}
  {\bibfield  {journal} {\bibinfo  {journal} {Water Resources Research}\
  }\textbf {\bibinfo {volume} {21}},\ \bibinfo {pages} {951 – 958} (\bibinfo
  {year} {1985})},\ \bibinfo {note} {cited by: 113}\BibitemShut {NoStop}%
\bibitem [{\citenamefont {Tufenkji}(2007)}]{Tufenkji2007}%
  \BibitemOpen
  \bibfield  {author} {\bibinfo {author} {\bibfnamefont {N.}~\bibnamefont
  {Tufenkji}},\ }\bibinfo {title} {Colloid and microbe migration in granular
  environments: A discussion of modelling methods},\ in\ \href
  {https://doi.org/10.1007/978-3-540-71339-5_5} {\emph {\bibinfo {booktitle}
  {Colloidal Transport in Porous Media}}},\ \bibinfo {editor} {edited by\
  \bibinfo {editor} {\bibfnamefont {F.~H.}\ \bibnamefont {Frimmel}}, \bibinfo
  {editor} {\bibfnamefont {F.}~\bibnamefont {Von Der~Kammer}},\ and\ \bibinfo
  {editor} {\bibfnamefont {H.-C.}\ \bibnamefont {Flemming}}}\ (\bibinfo
  {publisher} {Springer Berlin Heidelberg},\ \bibinfo {address} {Berlin,
  Heidelberg},\ \bibinfo {year} {2007})\ pp.\ \bibinfo {pages}
  {119--142}\BibitemShut {NoStop}%
\bibitem [{\citenamefont {de~Anna}\ \emph {et~al.}(2021)\citenamefont
  {de~Anna}, \citenamefont {Pahlavan}, \citenamefont {Yawata}, \citenamefont
  {Stocker},\ and\ \citenamefont {Juanes}}]{MicrobeNPAnna}%
  \BibitemOpen
  \bibfield  {author} {\bibinfo {author} {\bibfnamefont {P.}~\bibnamefont
  {de~Anna}}, \bibinfo {author} {\bibfnamefont {A.~A.}\ \bibnamefont
  {Pahlavan}}, \bibinfo {author} {\bibfnamefont {Y.}~\bibnamefont {Yawata}},
  \bibinfo {author} {\bibfnamefont {R.}~\bibnamefont {Stocker}},\ and\ \bibinfo
  {author} {\bibfnamefont {R.}~\bibnamefont {Juanes}},\ }\bibfield  {title}
  {\bibinfo {title} {Chemotaxis under flow disorder shapes microbial dispersion
  in porous media},\ }\href
  {https://doi.org/https://doi.org/10.1038/s41567-020-1002-x} {\bibfield
  {journal} {\bibinfo  {journal} {Nature Physics}\ }\textbf {\bibinfo {volume}
  {17}},\ \bibinfo {pages} {68} (\bibinfo {year} {2021})}\BibitemShut {NoStop}%
\bibitem [{\citenamefont {Bizmark}\ \emph {et~al.}(2020)\citenamefont
  {Bizmark}, \citenamefont {Schneider}, \citenamefont {Priestley},\ and\
  \citenamefont {Datta}}]{colloid_review}%
  \BibitemOpen
  \bibfield  {author} {\bibinfo {author} {\bibfnamefont {N.}~\bibnamefont
  {Bizmark}}, \bibinfo {author} {\bibfnamefont {J.}~\bibnamefont {Schneider}},
  \bibinfo {author} {\bibfnamefont {R.~D.}\ \bibnamefont {Priestley}},\ and\
  \bibinfo {author} {\bibfnamefont {S.~S.}\ \bibnamefont {Datta}},\ }\bibfield
  {title} {\bibinfo {title} {Multiscale dynamics of colloidal deposition and
  erosion in porous media},\ }\href {https://doi.org/10.1126/sciadv.abc2530}
  {\bibfield  {journal} {\bibinfo  {journal} {Science Advances}\ }\textbf
  {\bibinfo {volume} {6}},\ \bibinfo {pages} {eabc2530} (\bibinfo {year}
  {2020})}\BibitemShut {NoStop}%
\bibitem [{\citenamefont {Hewitt}(2020)}]{hewitt2020vigorous}%
  \BibitemOpen
  \bibfield  {author} {\bibinfo {author} {\bibfnamefont {D.~R.}\ \bibnamefont
  {Hewitt}},\ }\bibfield  {title} {\bibinfo {title} {Vigorous convection in
  porous media},\ }\href {https://doi.org/10.1098/rspa.2020.0111} {\bibfield
  {journal} {\bibinfo  {journal} {Proceedings of the Royal Society A:
  Mathematical, Physical and Engineering Sciences}\ }\textbf {\bibinfo {volume}
  {476}},\ \bibinfo {pages} {20200111} (\bibinfo {year} {2020})}\BibitemShut
  {NoStop}%
\bibitem [{\citenamefont {Pirozzoli}\ \emph {et~al.}(2021)\citenamefont
  {Pirozzoli}, \citenamefont {De~Paoli}, \citenamefont {Zonta},\ and\
  \citenamefont {Soldati}}]{pirozzoli2021ultimate}%
  \BibitemOpen
  \bibfield  {author} {\bibinfo {author} {\bibfnamefont {S.}~\bibnamefont
  {Pirozzoli}}, \bibinfo {author} {\bibfnamefont {M.}~\bibnamefont {De~Paoli}},
  \bibinfo {author} {\bibfnamefont {F.}~\bibnamefont {Zonta}},\ and\ \bibinfo
  {author} {\bibfnamefont {A.}~\bibnamefont {Soldati}},\ }\bibfield  {title}
  {\bibinfo {title} {Towards the ultimate regime in {{Rayleigh}}--{{Darcy}}
  convection},\ }\href {https://doi.org/10.1017/jfm.2020.1178} {\bibfield
  {journal} {\bibinfo  {journal} {Journal of Fluid Mechanics}\ }\textbf
  {\bibinfo {volume} {911}},\ \bibinfo {pages} {R4} (\bibinfo {year}
  {2021})}\BibitemShut {NoStop}%
\bibitem [{\citenamefont {De~Paoli}\ \emph {et~al.}(2022)\citenamefont
  {De~Paoli}, \citenamefont {Pirozzoli}, \citenamefont {Zonta},\ and\
  \citenamefont {Soldati}}]{depaoli2022strong}%
  \BibitemOpen
  \bibfield  {author} {\bibinfo {author} {\bibfnamefont {M.}~\bibnamefont
  {De~Paoli}}, \bibinfo {author} {\bibfnamefont {S.}~\bibnamefont {Pirozzoli}},
  \bibinfo {author} {\bibfnamefont {F.}~\bibnamefont {Zonta}},\ and\ \bibinfo
  {author} {\bibfnamefont {A.}~\bibnamefont {Soldati}},\ }\bibfield  {title}
  {\bibinfo {title} {Strong {{Rayleigh}}--{{Darcy}} convection regime in
  three-dimensional porous media},\ }\href
  {https://doi.org/10.1017/jfm.2022.461} {\bibfield  {journal} {\bibinfo
  {journal} {Journal of Fluid Mechanics}\ }\textbf {\bibinfo {volume} {943}},\
  \bibinfo {pages} {A51} (\bibinfo {year} {2022})}\BibitemShut {NoStop}%
\bibitem [{\citenamefont {Hidalgo}\ \emph {et~al.}(2012)\citenamefont
  {Hidalgo}, \citenamefont {Fe}, \citenamefont {Cueto-Felgueroso},\ and\
  \citenamefont {Juanes}}]{PhysRevLett.109.264503}%
  \BibitemOpen
  \bibfield  {author} {\bibinfo {author} {\bibfnamefont {J.~J.}\ \bibnamefont
  {Hidalgo}}, \bibinfo {author} {\bibfnamefont {J.}~\bibnamefont {Fe}},
  \bibinfo {author} {\bibfnamefont {L.}~\bibnamefont {Cueto-Felgueroso}},\ and\
  \bibinfo {author} {\bibfnamefont {R.}~\bibnamefont {Juanes}},\ }\bibfield
  {title} {\bibinfo {title} {Scaling of convective mixing in porous media},\
  }\href {https://doi.org/10.1103/PhysRevLett.109.264503} {\bibfield  {journal}
  {\bibinfo  {journal} {Phys. Rev. Lett.}\ }\textbf {\bibinfo {volume} {109}},\
  \bibinfo {pages} {264503} (\bibinfo {year} {2012})}\BibitemShut {NoStop}%
\bibitem [{\citenamefont {Hewitt}\ \emph {et~al.}(2012)\citenamefont {Hewitt},
  \citenamefont {Neufeld},\ and\ \citenamefont {Lister}}]{hewitt2012ultimate}%
  \BibitemOpen
  \bibfield  {author} {\bibinfo {author} {\bibfnamefont {D.~R.}\ \bibnamefont
  {Hewitt}}, \bibinfo {author} {\bibfnamefont {J.~A.}\ \bibnamefont
  {Neufeld}},\ and\ \bibinfo {author} {\bibfnamefont {J.~R.}\ \bibnamefont
  {Lister}},\ }\bibfield  {title} {\bibinfo {title} {Ultimate regime of high
  {{Rayleigh}} number convection in a porous medium},\ }\href
  {https://doi.org/10.1103/PhysRevLett.108.224503} {\bibfield  {journal}
  {\bibinfo  {journal} {Phys. Rev. Lett.}\ }\textbf {\bibinfo {volume} {108}},\
  \bibinfo {pages} {224503} (\bibinfo {year} {2012})}\BibitemShut {NoStop}%
\bibitem [{\citenamefont {Otero}\ \emph {et~al.}(2004)\citenamefont {Otero},
  \citenamefont {Dontcheva}, \citenamefont {Johnston}, \citenamefont
  {Worthing}, \citenamefont {Kurganov}, \citenamefont {Petrova},\ and\
  \citenamefont {Doering}}]{otero2004highrayleighnumber}%
  \BibitemOpen
  \bibfield  {author} {\bibinfo {author} {\bibfnamefont {J.}~\bibnamefont
  {Otero}}, \bibinfo {author} {\bibfnamefont {L.~A.}\ \bibnamefont
  {Dontcheva}}, \bibinfo {author} {\bibfnamefont {H.}~\bibnamefont {Johnston}},
  \bibinfo {author} {\bibfnamefont {R.~A.}\ \bibnamefont {Worthing}}, \bibinfo
  {author} {\bibfnamefont {A.}~\bibnamefont {Kurganov}}, \bibinfo {author}
  {\bibfnamefont {G.}~\bibnamefont {Petrova}},\ and\ \bibinfo {author}
  {\bibfnamefont {C.~R.}\ \bibnamefont {Doering}},\ }\bibfield  {title}
  {\bibinfo {title} {High-{{Rayleigh-number}} convection in a fluid-saturated
  porous layer},\ }\href {https://doi.org/10.1017/S0022112003007298} {\bibfield
   {journal} {\bibinfo  {journal} {Journal of Fluid Mechanics}\ }\textbf
  {\bibinfo {volume} {500}},\ \bibinfo {pages} {263} (\bibinfo {year}
  {2004})}\BibitemShut {NoStop}%
\bibitem [{\citenamefont {Slim}(2014)}]{slim2014solutalconvection}%
  \BibitemOpen
  \bibfield  {author} {\bibinfo {author} {\bibfnamefont {A.~C.}\ \bibnamefont
  {Slim}},\ }\bibfield  {title} {\bibinfo {title} {Solutal-convection regimes
  in a two-dimensional porous medium},\ }\href
  {https://doi.org/10.1017/jfm.2013.673} {\bibfield  {journal} {\bibinfo
  {journal} {Journal of Fluid Mechanics}\ }\textbf {\bibinfo {volume} {741}},\
  \bibinfo {pages} {461} (\bibinfo {year} {2014})}\BibitemShut {NoStop}%
\bibitem [{\citenamefont {Salibindla}\ \emph {et~al.}(2018)\citenamefont
  {Salibindla}, \citenamefont {Subedi}, \citenamefont {Shen}, \citenamefont
  {Masuk},\ and\ \citenamefont {Ni}}]{salibindla2018dissolutiondriven}%
  \BibitemOpen
  \bibfield  {author} {\bibinfo {author} {\bibfnamefont {A.~K.~R.}\
  \bibnamefont {Salibindla}}, \bibinfo {author} {\bibfnamefont
  {R.}~\bibnamefont {Subedi}}, \bibinfo {author} {\bibfnamefont {V.~C.}\
  \bibnamefont {Shen}}, \bibinfo {author} {\bibfnamefont {A.~U.~M.}\
  \bibnamefont {Masuk}},\ and\ \bibinfo {author} {\bibfnamefont
  {R.}~\bibnamefont {Ni}},\ }\bibfield  {title} {\bibinfo {title}
  {Dissolution-driven convection in a heterogeneous porous medium},\ }\href
  {https://doi.org/10.1017/jfm.2018.732} {\bibfield  {journal} {\bibinfo
  {journal} {Journal of Fluid Mechanics}\ }\textbf {\bibinfo {volume} {857}},\
  \bibinfo {pages} {61} (\bibinfo {year} {2018})}\BibitemShut {NoStop}%
\bibitem [{\citenamefont {Daniel}\ and\ \citenamefont
  {Riaz}(2014)}]{daniel2014effect}%
  \BibitemOpen
  \bibfield  {author} {\bibinfo {author} {\bibfnamefont {D.}~\bibnamefont
  {Daniel}}\ and\ \bibinfo {author} {\bibfnamefont {A.}~\bibnamefont {Riaz}},\
  }\bibfield  {title} {\bibinfo {title} {Effect of viscosity contrast on
  gravitationally unstable diffusive layers in porous media},\ }\href
  {https://doi.org/10.1063/1.4900843} {\bibfield  {journal} {\bibinfo
  {journal} {Physics of Fluids}\ }\textbf {\bibinfo {volume} {26}},\ \bibinfo
  {pages} {116601} (\bibinfo {year} {2014})}\BibitemShut {NoStop}%
\bibitem [{\citenamefont {Hidalgo}\ \emph {et~al.}(2015)\citenamefont
  {Hidalgo}, \citenamefont {Dentz}, \citenamefont {Cabeza},\ and\ \citenamefont
  {Carrera}}]{hidalgo2015dissolution}%
  \BibitemOpen
  \bibfield  {author} {\bibinfo {author} {\bibfnamefont {J.~J.}\ \bibnamefont
  {Hidalgo}}, \bibinfo {author} {\bibfnamefont {M.}~\bibnamefont {Dentz}},
  \bibinfo {author} {\bibfnamefont {Y.}~\bibnamefont {Cabeza}},\ and\ \bibinfo
  {author} {\bibfnamefont {J.}~\bibnamefont {Carrera}},\ }\bibfield  {title}
  {\bibinfo {title} {Dissolution patterns and mixing dynamics in unstable
  reactive flow},\ }\href {https://doi.org/10.1002/2015GL065036} {\bibfield
  {journal} {\bibinfo  {journal} {Geophysical Research Letters}\ }\textbf
  {\bibinfo {volume} {42}},\ \bibinfo {pages} {6357} (\bibinfo {year}
  {2015})}\BibitemShut {NoStop}%
\bibitem [{\citenamefont {Fu}\ \emph {et~al.}(2015)\citenamefont {Fu},
  \citenamefont {{Cueto-Felgueroso}}, \citenamefont {Bolster},\ and\
  \citenamefont {Juanes}}]{fu2015rock}%
  \BibitemOpen
  \bibfield  {author} {\bibinfo {author} {\bibfnamefont {X.}~\bibnamefont
  {Fu}}, \bibinfo {author} {\bibfnamefont {L.}~\bibnamefont
  {{Cueto-Felgueroso}}}, \bibinfo {author} {\bibfnamefont {D.}~\bibnamefont
  {Bolster}},\ and\ \bibinfo {author} {\bibfnamefont {R.}~\bibnamefont
  {Juanes}},\ }\bibfield  {title} {\bibinfo {title} {Rock dissolution patterns
  and geochemical shutdown of --brine--carbonate reactions during convective
  mixing in porous media},\ }\href {https://doi.org/10.1017/jfm.2014.647}
  {\bibfield  {journal} {\bibinfo  {journal} {Journal of Fluid Mechanics}\
  }\textbf {\bibinfo {volume} {764}},\ \bibinfo {pages} {296} (\bibinfo {year}
  {2015})}\BibitemShut {NoStop}%
\bibitem [{Note1()}]{Note1}%
  \BibitemOpen
  \bibinfo {note} {See Supplemental Material at [URL will be inserted by
  publisher] for numerical methods.}\BibitemShut {Stop}%
\bibitem [{\citenamefont {Hu}\ \emph {et~al.}(2023)\citenamefont {Hu},
  \citenamefont {Xu},\ and\ \citenamefont {Yang}}]{hu2023effects}%
  \BibitemOpen
  \bibfield  {author} {\bibinfo {author} {\bibfnamefont {C.}~\bibnamefont
  {Hu}}, \bibinfo {author} {\bibfnamefont {K.}~\bibnamefont {Xu}},\ and\
  \bibinfo {author} {\bibfnamefont {Y.}~\bibnamefont {Yang}},\ }\bibfield
  {title} {\bibinfo {title} {Effects of the geothermal gradient on the
  convective dissolution in {{CO$_2$}} sequestration},\ }\href
  {https://doi.org/10.1017/jfm.2023.349} {\bibfield  {journal} {\bibinfo
  {journal} {Journal of Fluid Mechanics}\ }\textbf {\bibinfo {volume} {963}},\
  \bibinfo {pages} {A23} (\bibinfo {year} {2023})}\BibitemShut {NoStop}%
\bibitem [{Note2()}]{Note2}%
  \BibitemOpen
  \bibinfo {note} {See Supplemental Material at [URL will be inserted by
  publisher] for a movie of temperature and concentration at
  $Ra=500$.}\BibitemShut {Stop}%
\bibitem [{Note3()}]{Note3}%
  \BibitemOpen
  \bibinfo {note} {See Supplemental Material at [URL will be inserted by
  publisher] for calculating the width of cell $l_c$.}\BibitemShut {Stop}%
\bibitem [{Note4()}]{Note4}%
  \BibitemOpen
  \bibinfo {note} {See Supplemental Material at [URL will be inserted by
  publisher] for a movie of temperature and concentration at $Ra=2\times
  10^4$.}\BibitemShut {Stop}%
\bibitem [{\citenamefont {Zhu}\ \emph {et~al.}(2024)\citenamefont {Zhu},
  \citenamefont {Fu},\ and\ \citenamefont {Paoli}}]{zhu2024transporta}%
  \BibitemOpen
  \bibfield  {author} {\bibinfo {author} {\bibfnamefont {X.}~\bibnamefont
  {Zhu}}, \bibinfo {author} {\bibfnamefont {Y.}~\bibnamefont {Fu}},\ and\
  \bibinfo {author} {\bibfnamefont {M.~D.}\ \bibnamefont {Paoli}},\ }\bibfield
  {title} {\bibinfo {title} {Transport scaling in porous media convection},\
  }\href {https://doi.org/10.1017/jfm.2024.528} {\bibfield  {journal} {\bibinfo
   {journal} {Journal of Fluid Mechanics}\ }\textbf {\bibinfo {volume} {991}},\
  \bibinfo {pages} {A4} (\bibinfo {year} {2024})}\BibitemShut {NoStop}%
\bibitem [{Note5()}]{Note5}%
  \BibitemOpen
  \bibinfo {note} {See Supplemental Material at [URL will be inserted by
  publisher] for the illustration of determining boundary layer
  thickness.}\BibitemShut {Stop}%
\end{thebibliography}%


\begin{thebibliography}{1}%
\makeatletter
\providecommand \@ifxundefined [1]{%
 \@ifx{#1\undefined}
}%
\providecommand \@ifnum [1]{%
 \ifnum #1\expandafter \@firstoftwo
 \else \expandafter \@secondoftwo
 \fi
}%
\providecommand \@ifx [1]{%
 \ifx #1\expandafter \@firstoftwo
 \else \expandafter \@secondoftwo
 \fi
}%
\providecommand \natexlab [1]{#1}%
\providecommand \enquote  [1]{``#1''}%
\providecommand \bibnamefont  [1]{#1}%
\providecommand \bibfnamefont [1]{#1}%
\providecommand \citenamefont [1]{#1}%
\providecommand \href@noop [0]{\@secondoftwo}%
\providecommand \href [0]{\begingroup \@sanitize@url \@href}%
\providecommand \@href[1]{\@@startlink{#1}\@@href}%
\providecommand \@@href[1]{\endgroup#1\@@endlink}%
\providecommand \@sanitize@url [0]{\catcode `\\12\catcode `\$12\catcode
  `\&12\catcode `\#12\catcode `\^12\catcode `\_12\catcode `\%12\relax}%
\providecommand \@@startlink[1]{}%
\providecommand \@@endlink[0]{}%
\providecommand \url  [0]{\begingroup\@sanitize@url \@url }%
\providecommand \@url [1]{\endgroup\@href {#1}{\urlprefix }}%
\providecommand \urlprefix  [0]{URL }%
\providecommand \Eprint [0]{\href }%
\providecommand \doibase [0]{https://doi.org/}%
\providecommand \selectlanguage [0]{\@gobble}%
\providecommand \bibinfo  [0]{\@secondoftwo}%
\providecommand \bibfield  [0]{\@secondoftwo}%
\providecommand \translation [1]{[#1]}%
\providecommand \BibitemOpen [0]{}%
\providecommand \bibitemStop [0]{}%
\providecommand \bibitemNoStop [0]{.\EOS\space}%
\providecommand \EOS [0]{\spacefactor3000\relax}%
\providecommand \BibitemShut  [1]{\csname bibitem#1\endcsname}%
\let\auto@bib@innerbib\@empty
\bibitem [{\citenamefont {Hu}\ \emph {et~al.}(2023)\citenamefont {Hu},
  \citenamefont {Xu},\ and\ \citenamefont {Yang}}]{hu2023effects}%
  \BibitemOpen
  \bibfield  {author} {\bibinfo {author} {\bibfnamefont {C.}~\bibnamefont
  {Hu}}, \bibinfo {author} {\bibfnamefont {K.}~\bibnamefont {Xu}},\ and\
  \bibinfo {author} {\bibfnamefont {Y.}~\bibnamefont {Yang}},\ }\bibfield
  {title} {\bibinfo {title} {Effects of the geothermal gradient on the
  convective dissolution in {{CO$_2$}} sequestration},\ }\href
  {https://doi.org/10.1017/jfm.2023.349} {\bibfield  {journal} {\bibinfo
  {journal} {Journal of Fluid Mechanics}\ }\textbf {\bibinfo {volume} {963}},\
  \bibinfo {pages} {A23} (\bibinfo {year} {2023})}\BibitemShut {NoStop}%
\end{thebibliography}%

\end{document}



\title{Supplementary Materials for ``Passive scalar dispersion along porous stratum with natural convection''}


\author{Chenglong Hu}
\affiliation{State Key Laboratory for Turbulence and Complex Systems, and Department of Mechanics and Engineering Science, College of Engineering, Peking University, Beijing 100871, P.R. China}

\author{Ke Xu}
\email{kexu1989@pku.edu.cn}
\affiliation{Department of Energy and Resources Engineering, College of Engineering, Peking University, Beijing 100871, P.R. China}

\author{Yantao Yang}
\email[]{yantao.yang@pku.edu.cn}
\affiliation{State Key Laboratory for Turbulence and Complex Systems, and Department of Mechanics and Engineering Science, College of Engineering, Peking University, Beijing 100871, P.R. China}
\affiliation{Laoshan Laboratory, Qingdao 266299, Shandong, P.R. China}


\date{\today}



\maketitle
\newpage 

\section{A. Theoretical formulation and numerical method}

Consider a two-dimensional fluid-saturated porous medium with constant porosity $\phi$ and permeability $K$, contained within a rectangular cell of height $H$ and width $L$. The medium is heated from below and cooled from above. The density of fluid decreases linearly with temperature and is independent of concentration, resulting in an unstable density difference $\Delta \rho_T$ across the domain. The incompressible flow in the porous medium is governed by Darcy's law, which describes the velocity $\boldsymbol{u}=(u,w)$, where $u$ and $w$ are the velocity components in the $x$ and $z$ directions, respectively. Meanwhile, the temperature and concentration fields are governed by the advection-diffusion equations with constant diffusivities $\kappa_T$ and $\kappa_S$, respectively. Here we assume that there are no heat or mass transfer between the porous medium and the fluid. By employing the characteristic velocity $U=Kg \Delta \rho_T/\mu$ ($\mu$ is the viscosity of the fluid), height $H$ and time scale $t_c=\phi H/U$, the non-dimensionalized governing equations read
\begin{subequations}\label{eqn:ges}
  \begin{eqnarray}
    \nabla \cdot \boldsymbol{u} &=& 0, \label{eqn:ges-con}\\
    \boldsymbol{u} &=& -\nabla P +T \boldsymbol{e}_z, \label{eqn:ges-vel}\\
    \frac{\partial T}{\partial t} &=& -\boldsymbol{u} \cdot \nabla T+\frac{1}{Ra_T} \nabla^{2} T, \label{eqn:ges-T}\\
    \frac{\partial S}{\partial t} &=& -\boldsymbol{u} \cdot \nabla S+\frac{1}{Le Ra_T} \nabla^{2} S, \label{eqn:ges-S}
  \end{eqnarray}
\end{subequations}
where $\boldsymbol{e}_z$ is the vertical unit vector in the direction opposite to gravity. The boundary conditions at the bottom and top are
\begin{subequations}\label{eqn:bcs}
  \begin{eqnarray}
    && w=0,~ T=1,~ \frac{\partial S}{\partial z}=0,~~\mbox{at}~ z=0, \\
    && w=0,~ T=0,~ \frac{\partial S}{\partial z}=0,~~\mbox{at}~ z=1.
  \end{eqnarray}
\end{subequations}
Horizontal periodicity is employed for all variables. There are two controlling parameters in the equations: the Rayleigh number $Ra=KHg \Delta \rho_T /\phi\mu\kappa_T$ and the Lewis number $Le=\kappa_T/\kappa_S$. The numerical methods of solving the above equations are reported in detail in our previous study \cite{hu2023effects}. We begin by taking the divergence of the equation \eqref{eqn:ges-vel} and utilizing the continuity equation \eqref{eqn:ges-con} for incompressible flow. This leads to a Poisson equation for pressure as
\begin{equation}\label{eqn:poipr}
    \nabla^2 p=\frac{\partial T}{\partial z},
\end{equation}
where the boundary conditions at the bottom and top plates are $\displaystyle\left.\frac{\partial p}{\partial z}\right|_{z=0,1}=T|_{z=0,1}$. Due to horizontal periodicity, a Fourier transform is applied in the $x$ direction. The Poisson equation \eqref{eqn:poipr} can then be numerically solved by solving a set of tridiagonal systems in the $z$ direction. The velocity is readily to be calculated by the equation \eqref{eqn:ges-vel}. The treatment of the advection-diffusion equation is similar for both temperature and concentration. The advection and diffusion terms are treated explicitly and implicitly in time, respectively. A low-storage third-order Runge-Kutta scheme is used for time advancement. All variables are discretized by a second-order finite-difference method on staggered grids.

\section{B. Details of numerical settings for the concentration field}
We first conduct simulations for equations \eqref{eqn:ges-con}-\eqref{eqn:ges-T} to a statistically steady state, which is taken as the initial condition for temperature and velocity in the next step. Then the concentration field is introduced into the flow with an initial distribution as $S_0(x)=\exp\left[-(x-x_0)^2/(4\tau)\right]$, where $x_0$ and $\tau$ are two parameters that determine the center and shape of the initial concentration profile, respectively. In our simulations, the Rayleigh number ranges from $10$ to $2\times10^4$ and the Lewis number ranges from $1$ to $4$. The width of the computational domain $L/H$ varies with $Ra$ to ensure approximately twenty pairs of convection rolls along the horizontal direction for statistical convergence, which varies from $L/H=40$ for $Ra=10$ to $L/H=5$ for $Ra=2\times10^4$. The center location $x_0$ is set at five equidistant positions $\{0.1L/H, 0.3L/H, 0.5L/H, 0.7L/H, 0.9L/H\}$. The parameter $\tau$ is determined to satisfy $S_0(x_0 \pm 0.15L/H)=0.01$, which means that the initial area of concentration greater than 0.01 covers $30\%$ of the domain width. The effective dispersion coefficient is calculated every unit time by fitting the vertically averaged concentration profiles with the analytical solution for 1D diffusion equation. In Tables~\ref{tab:regime1} to \ref{tab:regime3_Le4}, we summarize the details of numerical settings with the averaged statistics obtained from simulations for three regimes depicted in Fig.~2(b) in the main context, respectively.


\section{C. Calculations of the characteristic length for horizontal dispersion}
\begin{figure}
  \centering
  \subfigure[]{\includegraphics[width=0.6\textwidth]{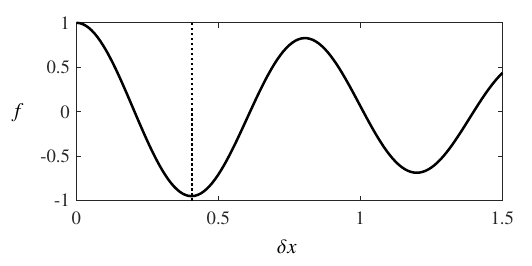}}
  \subfigure[]{\includegraphics[width=0.25\textwidth]{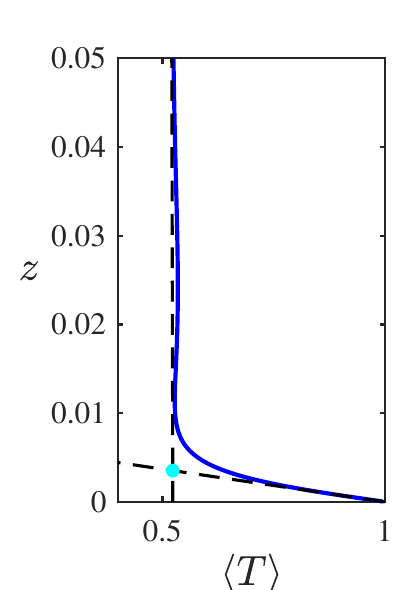}}
  \caption{\label{fig:length} Determinations of the characteristic length scales. (a) The autocorrelation function $f(\delta x)$ of temperature at mid-height for $Ra=500$. The dashed line denotes the location of the first minimum of $f(\delta x)$, i.e. the value of $l_c$. (b) The thermal boundary thickness $\delta$ for $Ra=2\times10^4$. The dashed lines denote linear fits of the temperature mean profile in two regions.}
\end{figure}

In the main context, we extract different length scales as the characteristic lengths to measure the horizontal dispersion coefficient. For the cell-dominated regime, the characteristic length is defined as the width of convection cells $l_c$, which can be obtained from the autocorrelation function $f(\delta x)$ of the temperature filed at mid-height. The value of $l_c$ corresponds to the location of the first minimum of $f(\delta x)$ as shown in Fig.~\ref{fig:length}(a). For the plume-dominated regime, the thickness of thermal boundary layer $\delta$ is chosen as the characteristic length, which is determined by the intersection between two linearly fitted lines of temperature profile in the interior ($0.4\le z/H \le 0.6$) and near-wall regions (ten grids closest to the boundaries), as shown in Fig.~\ref{fig:length}(b).

\squeezetable
\begin{table}
\caption{\label{tab:regime1} Simulation details for regime \uppercase\expandafter{\romannumeral1} (diffusion regime) at $Le=2$. The columns from left to right represent the Rayleigh number $Ra$, the non-dimensionalized domain width $L/H$, the number of grids in the $x$ direction $N_x$, the number of grids in the $z$ direction $N_z$, the non-dimensionalized time step $\Delta t$, the non-dimensionalized total simulation time $N_t$, the center location of the initial concentration profile $x_0$, the parameter $\tau$ to determine the width of the initial concentration profile, the Nusselt number $Nu$, and the fitted effective dispersion coefficient $\hat{D}/D_0$ averaged over the last 100 time period, respectively}
  \begin{ruledtabular}
    \begin{tabular}{ccccccccccc}
        $Ra$ & $Le$ & $L/H$ & $N_x$ & $N_z$ & $\Delta t$ & $N_t$ & $x_0$ & $\tau$ & $Nu$ & $\hat{D}/D_0$ \\
        \colrule
        \multirow{5}{*}{$10$} & \multirow{5}{*}{$2$} & \multirow{5}{*}{$40$} & \multirow{5}{*}{$1920$} & \multirow{5}{*}{$96$} & \multirow{5}{*}{1.0E-01} & \multirow{5}{*}{$400$} & $0.1L/H$ & \multirow{5}{*}{$1.95$} & \multirow{5}{*}{1.00} & 1.00 \\
         &  &  &  &  &  &  & $0.3L/H$ &  &  & 1.00 \\
         &  &  &  &  &  &  & $0.5L/H$ &  &  & 1.00 \\
         &  &  &  &  &  &  & $0.7L/H$ &  &  & 1.00 \\
         &  &  &  &  &  &  & $0.9L/H$ &  &  & 1.00 \\
         \colrule
        \multirow{5}{*}{$20$} & \multirow{5}{*}{$2$} & \multirow{5}{*}{$40$} & \multirow{5}{*}{$1920$} & \multirow{5}{*}{$96$} & \multirow{5}{*}{1.0E-01} & \multirow{5}{*}{$400$} & $0.1L/H$ & \multirow{5}{*}{$1.95$} & \multirow{5}{*}{1.00} & 1.00 \\
         &  &  &  &  &  &  & $0.3L/H$ &  &  & 1.00 \\
         &  &  &  &  &  &  & $0.5L/H$ &  &  & 1.00 \\
         &  &  &  &  &  &  & $0.7L/H$ &  &  & 1.00 \\
         &  &  &  &  &  &  & $0.9L/H$ &  &  & 1.00 \\
         \colrule
        \multirow{5}{*}{$30$} & \multirow{5}{*}{$2$} & \multirow{5}{*}{$40$} & \multirow{5}{*}{$1920$} & \multirow{5}{*}{$96$} & \multirow{5}{*}{1.0E-01} & \multirow{5}{*}{$400$} & $0.1L/H$ & \multirow{5}{*}{$1.95$} & \multirow{5}{*}{1.00} & 1.00 \\
         &  &  &  &  &  &  & $0.3L/H$ &  &  & 1.00 \\
         &  &  &  &  &  &  & $0.5L/H$ &  &  & 1.00 \\
         &  &  &  &  &  &  & $0.7L/H$ &  &  & 1.00 \\
         &  &  &  &  &  &  & $0.9L/H$ &  &  & 1.00
        \end{tabular}
  \end{ruledtabular}
\end{table}

\squeezetable
\begin{table}
  \caption{\label{tab:regime2} Simulation details for regime \uppercase\expandafter{\romannumeral2} (cell-dominated regime) at $Le=2$. The notations are consistent with those in Table~\ref{tab:regime1}. Additionally, $w^\mathit{rms}/U$ and $l_c$ represent the r.m.s. vertical velocity at mid-height and the width of the convection cell, respectively.}
  \begin{ruledtabular}
    \begin{tabular}{ccccccccccccc}
        $Ra$ & $Le$ & $L/H$ & $N_x$ & $N_z$ & $\Delta t$ & $N_t$ & $x_0$ & $\tau$ & $Nu$ & $w^\mathit{rms}/U$ & $\hat{D}/D_0$ & $l_c$ \\
        \colrule
        \multirow{5}{*}{$40$} & \multirow{5}{*}{$2$} & \multirow{5}{*}{$40$} & \multirow{5}{*}{$1920$} & \multirow{5}{*}{$96$} & \multirow{5}{*}{1.0E-01} & \multirow{5}{*}{$400$} & $0.1L/H$ & \multirow{5}{*}{$1.95$} & \multirow{5}{*}{1.03} & \multirow{5}{*}{0.025} & 1.10 & \multirow{5}{*}{1.000} \\
         &  &  &  &  &  &  & $0.3L/H$ &  &  &  & 1.10 &  \\
         &  &  &  &  &  &  & $0.5L/H$ &  &  &  & 1.10 &  \\
         &  &  &  &  &  &  & $0.7L/H$ &  &  &  & 1.10 &  \\
         &  &  &  &  &  &  & $0.9L/H$ &  &  &  & 1.10 &  \\
        \colrule
        \multirow{5}{*}{$50$} & \multirow{5}{*}{$2$} & \multirow{5}{*}{$40$} & \multirow{5}{*}{$1920$} & \multirow{5}{*}{$96$} & \multirow{5}{*}{5.0E-02} & \multirow{5}{*}{$400$} & $0.1L/H$ & \multirow{5}{*}{$1.95$} & \multirow{5}{*}{1.44} & \multirow{5}{*}{0.097} & 1.94 & \multirow{5}{*}{0.917} \\
         &  &  &  &  &  &  & $0.3L/H$ &  &  &  & 1.95 &  \\
         &  &  &  &  &  &  & $0.5L/H$ &  &  &  & 2.04 &  \\
         &  &  &  &  &  &  & $0.7L/H$ &  &  &  & 2.07 &  \\
         &  &  &  &  &  &  & $0.9L/H$ &  &  &  & 2.02 &  \\
        \colrule
        \multirow{5}{*}{$70$} & \multirow{5}{*}{$2$} & \multirow{5}{*}{$40$} & \multirow{5}{*}{$1920$} & \multirow{5}{*}{$96$} & \multirow{5}{*}{5.0E-02} & \multirow{5}{*}{$400$} & $0.1L/H$ & \multirow{5}{*}{$1.95$} & \multirow{5}{*}{1.98} & \multirow{5}{*}{0.128} & 2.53 & \multirow{5}{*}{0.771} \\
         &  &  &  &  &  &  & $0.3L/H$ &  &  &  & 2.45 &  \\
         &  &  &  &  &  &  & $0.5L/H$ &  &  &  & 2.49 &  \\
         &  &  &  &  &  &  & $0.7L/H$ &  &  &  & 2.52 &  \\
         &  &  &  &  &  &  & $0.9L/H$ &  &  &  & 2.50 &  \\
        \colrule
        \multirow{5}{*}{$100$} & \multirow{5}{*}{$2$} & \multirow{5}{*}{$30$} & \multirow{5}{*}{$1440$} & \multirow{5}{*}{$96$} & \multirow{5}{*}{5.0E-02} & \multirow{5}{*}{$400$} & $0.1L/H$ & \multirow{5}{*}{$1.10$} & \multirow{5}{*}{2.56} & \multirow{5}{*}{0.138} & 2.68 & \multirow{5}{*}{0.646} \\
         &  &  &  &  &  &  & $0.3L/H$ &  &  &  & 2.92 &  \\
         &  &  &  &  &  &  & $0.5L/H$ &  &  &  & 2.84 &  \\
         &  &  &  &  &  &  & $0.7L/H$ &  &  &  & 2.75 &  \\
         &  &  &  &  &  &  & $0.9L/H$ &  &  &  & 2.69 &  \\
        \colrule
        \multirow{5}{*}{$250$} & \multirow{5}{*}{$2$} & \multirow{5}{*}{$20$} & \multirow{5}{*}{$1440$} & \multirow{5}{*}{$144$} & \multirow{5}{*}{2.0E-02} & \multirow{5}{*}{$400$} & $0.1L/H$ & \multirow{5}{*}{$0.49$} & \multirow{5}{*}{4.58} & \multirow{5}{*}{0.130} & 3.60 & \multirow{5}{*}{0.472} \\
         &  &  &  &  &  &  & $0.3L/H$ &  &  &  & 3.50 &  \\
         &  &  &  &  &  &  & $0.5L/H$ &  &  &  & 3.42 &  \\
         &  &  &  &  &  &  & $0.7L/H$ &  &  &  & 3.47 &  \\
         &  &  &  &  &  &  & $0.9L/H$ &  &  &  & 3.36 &  \\
        \colrule
        \multirow{5}{*}{$500$} & \multirow{5}{*}{$2$} & \multirow{5}{*}{$18$} & \multirow{5}{*}{$1728$} & \multirow{5}{*}{$144$} & \multirow{5}{*}{2.0E-02} & \multirow{5}{*}{$400$} & $0.1L/H$ & \multirow{5}{*}{$0.40$} & \multirow{5}{*}{6.54} & \multirow{5}{*}{0.112} & 4.02 & \multirow{5}{*}{0.405} \\
         &  &  &  &  &  &  & $0.3L/H$ &  &  &  & 3.87 &  \\
         &  &  &  &  &  &  & $0.5L/H$ &  &  &  & 3.96 &  \\
         &  &  &  &  &  &  & $0.7L/H$ &  &  &  & 4.00 &  \\
         &  &  &  &  &  &  & $0.9L/H$ &  &  &  & 4.40 &  \\
        \colrule
        \multirow{5}{*}{$1000$} & \multirow{5}{*}{$2$} & \multirow{5}{*}{$15$} & \multirow{5}{*}{$2880$} & \multirow{5}{*}{$144$} & \multirow{5}{*}{1.0E-02} & \multirow{5}{*}{$400$} & $0.1L/H$ & \multirow{5}{*}{$0.27$} & \multirow{5}{*}{9.32} & \multirow{5}{*}{0.096} & 4.11 & \multirow{5}{*}{0.313} \\
         &  &  &  &  &  &  & $0.3L/H$ &  &  &  & 4.32 &  \\
         &  &  &  &  &  &  & $0.5L/H$ &  &  &  & 4.57 &  \\
         &  &  &  &  &  &  & $0.7L/H$ &  &  &  & 4.57 &  \\
         &  &  &  &  &  &  & $0.9L/H$ &  &  &  & 4.11 & 
        \end{tabular}
  \end{ruledtabular}
\end{table}

\begin{table}
  \caption{\label{tab:regime3} Simulation details for regime \uppercase\expandafter{\romannumeral3} (plume-dominated regime) at $Le=2$. The notations are consistent with those in Table~\ref{tab:regime1}. Additionally, $\delta$ and $u^\mathit{rms}/U$ represent the thermal boundary layer thickness and the r.m.s. horizontal velocity within the boundaries, respectively.}
    \begin{ruledtabular}
         \begin{tabular}{ccccccccccccc}
        $Ra$ & $Le$ & $L/H$ & $N_x$ & $N_z$ & $\Delta t$ & $N_t$ & $x_0$ & $\tau$ & $Nu$ & $u^\mathit{rms}/U$ & $\hat{D}/D_0$ & $\delta$ \\
        \colrule
        \multirow{5}{*}{$1350$} & \multirow{5}{*}{$2$} & \multirow{5}{*}{$15$} & \multirow{5}{*}{$3600$} & \multirow{5}{*}{$144$} & \multirow{5}{*}{1.0E-02} & \multirow{5}{*}{$400$} & $0.1L/H$ & \multirow{5}{*}{$0.27$} & \multirow{5}{*}{11.73} & \multirow{5}{*}{0.073} & 5.07 & \multirow{5}{*}{4.75E-02} \\
         &  &  &  &  &  &  & $0.3L/H$ &  &  &  & 5.84 &  \\
         &  &  &  &  &  &  & $0.5L/H$ &  &  &  & 7.06 &  \\
         &  &  &  &  &  &  & $0.7L/H$ &  &  &  & 5.97 &  \\
         &  &  &  &  &  &  & $0.9L/H$ &  &  &  & 5.06 &  \\
        \colrule
        \multirow{5}{*}{$1850$} & \multirow{5}{*}{$2$} & \multirow{5}{*}{$14$} & \multirow{5}{*}{$4032$} & \multirow{5}{*}{$192$} & \multirow{5}{*}{8.0E-03} & \multirow{5}{*}{$400$} & $0.1L/H$ & \multirow{5}{*}{$0.24$} & \multirow{5}{*}{15.23} & \multirow{5}{*}{0.078} & 7.01 & \multirow{5}{*}{3.32E-02} \\
         &  &  &  &  &  &  & $0.3L/H$ &  &  &  & 7.29 &  \\
         &  &  &  &  &  &  & $0.5L/H$ &  &  &  & 7.93 &  \\
         &  &  &  &  &  &  & $0.7L/H$ &  &  &  & 6.37 &  \\
         &  &  &  &  &  &  & $0.9L/H$ &  &  &  & 6.41 &  \\
        \colrule
        \multirow{5}{*}{$2500$} & \multirow{5}{*}{$2$} & \multirow{5}{*}{$12$} & \multirow{5}{*}{$4608$} & \multirow{5}{*}{$192$} & \multirow{5}{*}{5.0E-03} & \multirow{5}{*}{$400$} & $0.1L/H$ & \multirow{5}{*}{$0.18$} & \multirow{5}{*}{19.89} & \multirow{5}{*}{0.082} & 9.71 & \multirow{5}{*}{2.50E-02} \\
         &  &  &  &  &  &  & $0.3L/H$ &  &  &  & 9.91 &  \\
         &  &  &  &  &  &  & $0.5L/H$ &  &  &  & 9.73 &  \\
         &  &  &  &  &  &  & $0.7L/H$ &  &  &  & 7.98 &  \\
         &  &  &  &  &  &  & $0.9L/H$ &  &  &  & 9.12 &  \\
        \colrule
        \multirow{5}{*}{$5000$} & \multirow{5}{*}{$2$} & \multirow{5}{*}{$10$} & \multirow{5}{*}{$7680$} & \multirow{5}{*}{$288$} & \multirow{5}{*}{2.5E-03} & \multirow{5}{*}{$400$} & $0.1L/H$ & \multirow{5}{*}{$0.12$} & \multirow{5}{*}{37.24} & \multirow{5}{*}{0.085} & 10.87 & \multirow{5}{*}{1.32E-02} \\
         &  &  &  &  &  &  & $0.3L/H$ &  &  &  & 10.51 &  \\
         &  &  &  &  &  &  & $0.5L/H$ &  &  &  & 9.74 &  \\
         &  &  &  &  &  &  & $0.7L/H$ &  &  &  & 10.11 &  \\
         &  &  &  &  &  &  & $0.9L/H$ &  &  &  & 11.30 &  \\
        \colrule
        \multirow{5}{*}{$10000$} & \multirow{5}{*}{$2$} & \multirow{5}{*}{$6$} & \multirow{5}{*}{$9216$} & \multirow{5}{*}{$576$} & \multirow{5}{*}{1.0E-03} & \multirow{5}{*}{$400$} & $0.1L/H$ & \multirow{5}{*}{$0.04$} & \multirow{5}{*}{71.11} & \multirow{5}{*}{0.087} & 12.16 & \multirow{5}{*}{6.74E-03} \\
         &  &  &  &  &  &  & $0.3L/H$ &  &  &  & 10.69 &  \\
         &  &  &  &  &  &  & $0.5L/H$ &  &  &  & 9.59 &  \\
         &  &  &  &  &  &  & $0.7L/H$ &  &  &  & 10.36 &  \\
         &  &  &  &  &  &  & $0.9L/H$ &  &  &  & 10.85 &  \\
        \colrule
        \multirow{5}{*}{$20000$} & \multirow{5}{*}{$2$} & \multirow{5}{*}{$5$} & \multirow{5}{*}{$15360$} & \multirow{5}{*}{$1152$} & \multirow{5}{*}{5.0E-04} & \multirow{5}{*}{$200$} & $0.1L/H$ & \multirow{5}{*}{$0.03$} & \multirow{5}{*}{138.93} & \multirow{5}{*}{0.088} & 10.47 & \multirow{5}{*}{3.53E-03} \\
         &  &  &  &  &  &  & $0.3L/H$ &  &  &  & 10.71 &  \\
         &  &  &  &  &  &  & $0.5L/H$ &  &  &  & 10.30 &  \\
         &  &  &  &  &  &  & $0.7L/H$ &  &  &  & 10.25 &  \\
         &  &  &  &  &  &  & $0.9L/H$ &  &  &  & 9.41 & 
        \end{tabular}
    \end{ruledtabular}
\end{table}

\begin{table}
  \caption{\label{tab:regime3_Le1} Simulation details for regime \uppercase\expandafter{\romannumeral3} (plume-dominated regime) at $Le=1$. The notations are consistent with those in Table~\ref{tab:regime3}.}
    \begin{ruledtabular}
        \begin{tabular}{ccccccccccccc}
        $Ra$ & $Le$ & $L/H$ & $N_x$ & $N_z$ & $\Delta t$ & $N_t$ & $x_0$ & $\tau$ & $Nu$ & $u^\mathit{rms}/U$ & $\hat{D}/D_0$ & $\delta$ \\
        \colrule
        \multirow{5}{*}{$2500$} & \multirow{5}{*}{$1$} & \multirow{5}{*}{$12$} & \multirow{5}{*}{$4608$} & \multirow{5}{*}{$192$} & \multirow{5}{*}{5.0E-03} & \multirow{5}{*}{$400$} & $0.1L/H$ & \multirow{5}{*}{$0.18$} & \multirow{5}{*}{19.91} & \multirow{5}{*}{0.081} & 5.53 & \multirow{5}{*}{2.51E-02} \\
         &  &  &  &  &  &  & $0.3L/H$ &  &  &  & 5.77 &  \\
         &  &  &  &  &  &  & $0.5L/H$ &  &  &  & 5.68 &  \\
         &  &  &  &  &  &  & $0.7L/H$ &  &  &  & 4.63 &  \\
         &  &  &  &  &  &  & $0.9L/H$ &  &  &  & 5.19 &  \\
        \colrule
        \multirow{5}{*}{$5000$} & \multirow{5}{*}{$1$} & \multirow{5}{*}{$10$} & \multirow{5}{*}{$7680$} & \multirow{5}{*}{$288$} & \multirow{5}{*}{2.5E-03} & \multirow{5}{*}{$400$} & $0.1L/H$ & \multirow{5}{*}{$0.12$} & \multirow{5}{*}{37.19} & \multirow{5}{*}{0.084} & 6.67 & \multirow{5}{*}{1.33E-02} \\
         &  &  &  &  &  &  & $0.3L/H$ &  &  &  & 6.49 &  \\
         &  &  &  &  &  &  & $0.5L/H$ &  &  &  & 5.83 &  \\
         &  &  &  &  &  &  & $0.7L/H$ &  &  &  & 5.76 &  \\
         &  &  &  &  &  &  & $0.9L/H$ &  &  &  & 6.48 &  \\
        \colrule
        \multirow{5}{*}{$10000$} & \multirow{5}{*}{$1$} & \multirow{5}{*}{$6$} & \multirow{5}{*}{$9216$} & \multirow{5}{*}{$576$} & \multirow{5}{*}{1.0E-03} & \multirow{5}{*}{$400$} & $0.1L/H$ & \multirow{5}{*}{$0.04$} & \multirow{5}{*}{71.09} & \multirow{5}{*}{0.087} & 7.33 & \multirow{5}{*}{6.74E-03} \\
         &  &  &  &  &  &  & $0.3L/H$ &  &  &  & 7.03 &  \\
         &  &  &  &  &  &  & $0.5L/H$ &  &  &  & 6.88 &  \\
         &  &  &  &  &  &  & $0.7L/H$ &  &  &  & 7.08 &  \\
         &  &  &  &  &  &  & $0.9L/H$ &  &  &  & 7.57 &  \\
        \colrule
        \multirow{5}{*}{$20000$} & \multirow{5}{*}{$1$} & \multirow{5}{*}{$5$} & \multirow{5}{*}{$7680$} & \multirow{5}{*}{$576$} & \multirow{5}{*}{8.0E-04} & \multirow{5}{*}{$200$} & $0.1L/H$ & \multirow{5}{*}{$0.03$} & \multirow{5}{*}{141.22} & \multirow{5}{*}{0.089} & 6.68 & \multirow{5}{*}{3.54E-03} \\
         &  &  &  &  &  &  & $0.3L/H$ &  &  &  & 6.47 &  \\
         &  &  &  &  &  &  & $0.5L/H$ &  &  &  & 6.52 &  \\
         &  &  &  &  &  &  & $0.7L/H$ &  &  &  & 6.34 &  \\
         &  &  &  &  &  &  & $0.9L/H$ &  &  &  & 7.17 & 
        \end{tabular}
    \end{ruledtabular}
\end{table}

\begin{table}
  \caption{\label{tab:regime3_Le3} Simulation details for regime \uppercase\expandafter{\romannumeral3} (plume-dominated regime) at $Le=3$. The notations are consistent with those in Table~\ref{tab:regime3}.}
    \begin{ruledtabular}
        \begin{tabular}{ccccccccccccc}
        $Ra$ & $Le$ & $L/H$ & $N_x$ & $N_z$ & $\Delta t$ & $N_t$ & $x_0$ & $\tau$ & $Nu$ & $u^\mathit{rms}/U$ & $\hat{D}/D_0$ & $\delta$ \\
        \colrule
        \multirow{5}{*}{$2500$} & \multirow{5}{*}{$3$} & \multirow{5}{*}{$12$} & \multirow{5}{*}{$9216$} & \multirow{5}{*}{$288$} & \multirow{5}{*}{2.5E-03} & \multirow{5}{*}{$400$} & $0.1L/H$ & \multirow{5}{*}{$0.18$} & \multirow{5}{*}{19.92} & \multirow{5}{*}{0.082} & 11.54 & \multirow{5}{*}{2.40E-02} \\
         &  &  &  &  &  &  & $0.3L/H$ &  &  &  & 11.00 &  \\
         &  &  &  &  &  &  & $0.5L/H$ &  &  &  & 11.50 &  \\
         &  &  &  &  &  &  & $0.7L/H$ &  &  &  & 12.29 &  \\
         &  &  &  &  &  &  & $0.9L/H$ &  &  &  & 12.75 &  \\
         \colrule
        \multirow{5}{*}{$5000$} & \multirow{5}{*}{$3$} & \multirow{5}{*}{$10$} & \multirow{5}{*}{$15360$} & \multirow{5}{*}{$384$} & \multirow{5}{*}{2.0E-03} & \multirow{5}{*}{$400$} & $0.1L/H$ & \multirow{5}{*}{$0.12$} & \multirow{5}{*}{37.26} & \multirow{5}{*}{0.084} & 12.80 & \multirow{5}{*}{1.28E-02} \\
         &  &  &  &  &  &  & $0.3L/H$ &  &  &  & 13.48 &  \\
         &  &  &  &  &  &  & $0.5L/H$ &  &  &  & 14.60 &  \\
         &  &  &  &  &  &  & $0.7L/H$ &  &  &  & 14.67 &  \\
         &  &  &  &  &  &  & $0.9L/H$ &  &  &  & 14.00 &  \\
         \colrule
        \multirow{5}{*}{$10000$} & \multirow{5}{*}{$3$} & \multirow{5}{*}{$6$} & \multirow{5}{*}{$13824$} & \multirow{5}{*}{$768$} & \multirow{5}{*}{8.0E-04} & \multirow{5}{*}{$400$} & $0.1L/H$ & \multirow{5}{*}{$0.04$} & \multirow{5}{*}{70.78} & \multirow{5}{*}{0.087} & 15.62 & \multirow{5}{*}{6.68E-03} \\
         &  &  &  &  &  &  & $0.3L/H$ &  &  &  & 15.91 &  \\
         &  &  &  &  &  &  & $0.5L/H$ &  &  &  & 15.69 &  \\
         &  &  &  &  &  &  & $0.7L/H$ &  &  &  & 15.92 &  \\
         &  &  &  &  &  &  & $0.9L/H$ &  &  &  & 15.07 &  \\
         \colrule
        \multirow{5}{*}{$20000$} & \multirow{5}{*}{$3$} & \multirow{5}{*}{$5$} & \multirow{5}{*}{$15360$} & \multirow{5}{*}{$1152$} & \multirow{5}{*}{5.0E-04} & \multirow{5}{*}{$200$} & $0.1L/H$ & \multirow{5}{*}{$0.03$} & \multirow{5}{*}{139.30} & \multirow{5}{*}{0.089} & 14.02 & \multirow{5}{*}{3.52E-03} \\
         &  &  &  &  &  &  & $0.3L/H$ &  &  &  & 13.31 &  \\
         &  &  &  &  &  &  & $0.5L/H$ &  &  &  & 15.91 &  \\
         &  &  &  &  &  &  & $0.7L/H$ &  &  &  & 13.89 &  \\
         &  &  &  &  &  &  & $0.9L/H$ &  &  &  & 17.62 & 
        \end{tabular}
    \end{ruledtabular}
\end{table}

\begin{table}
  \caption{\label{tab:regime3_Le4} Simulation details for regime \uppercase\expandafter{\romannumeral3} (plume-dominated regime) at $Le=4$. The notations are consistent with those in Table~\ref{tab:regime3}.}
    \begin{ruledtabular}
        \begin{tabular}{ccccccccccccc}
        $Ra$ & $Le$ & $L/H$ & $N_x$ & $N_z$ & $\Delta t$ & $N_t$ & $x_0$ & $\tau$ & $Nu$ & $u^\mathit{rms}/U$ & $\hat{D}/D_0$ & $\delta$ \\
        \colrule
        \multirow{5}{*}{$2500$} & \multirow{5}{*}{$4$} & \multirow{5}{*}{$12$} & \multirow{5}{*}{$9216$} & \multirow{5}{*}{$288$} & \multirow{5}{*}{2.5E-03} & \multirow{5}{*}{$400$} & $0.1L/H$ & \multirow{5}{*}{$0.18$} & \multirow{5}{*}{19.92} & \multirow{5}{*}{0.082} & 14.28 & \multirow{5}{*}{2.40E-02} \\
         &  &  &  &  &  &  & $0.3L/H$ &  &  &  & 13.58 &  \\
         &  &  &  &  &  &  & $0.5L/H$ &  &  &  & 14.28 &  \\
         &  &  &  &  &  &  & $0.7L/H$ &  &  &  & 15.17 &  \\
         &  &  &  &  &  &  & $0.9L/H$ &  &  &  & 15.87 &  \\
        \colrule
        \multirow{5}{*}{$5000$} & \multirow{5}{*}{$4$} & \multirow{5}{*}{$10$} & \multirow{5}{*}{$15360$} & \multirow{5}{*}{$384$} & \multirow{5}{*}{2.0E-03} & \multirow{5}{*}{$400$} & $0.1L/H$ & \multirow{5}{*}{$0.12$} & \multirow{5}{*}{37.26} & \multirow{5}{*}{0.084} & 15.92 & \multirow{5}{*}{1.28E-02} \\
         &  &  &  &  &  &  & $0.3L/H$ &  &  &  & 16.79 &  \\
         &  &  &  &  &  &  & $0.5L/H$ &  &  &  & 18.36 &  \\
         &  &  &  &  &  &  & $0.7L/H$ &  &  &  & 18.46 &  \\
         &  &  &  &  &  &  & $0.9L/H$ &  &  &  & 17.49 &  \\
        \colrule
        \multirow{5}{*}{$10000$} & \multirow{5}{*}{$4$} & \multirow{5}{*}{$6$} & \multirow{5}{*}{$13824$} & \multirow{5}{*}{$768$} & \multirow{5}{*}{8.0E-04} & \multirow{5}{*}{$400$} & $0.1L/H$ & \multirow{5}{*}{$0.04$} & \multirow{5}{*}{70.78} & \multirow{5}{*}{0.087} & 19.69 & \multirow{5}{*}{6.68E-03} \\
         &  &  &  &  &  &  & $0.3L/H$ &  &  &  & 20.06 &  \\
         &  &  &  &  &  &  & $0.5L/H$ &  &  &  & 19.62 &  \\
         &  &  &  &  &  &  & $0.7L/H$ &  &  &  & 19.97 &  \\
         &  &  &  &  &  &  & $0.9L/H$ &  &  &  & 18.88 &  \\
        \colrule
        \multirow{5}{*}{$20000$} & \multirow{5}{*}{$4$} & \multirow{5}{*}{$5$} & \multirow{5}{*}{$15360$} & \multirow{5}{*}{$1152$} & \multirow{5}{*}{5.0E-04} & \multirow{5}{*}{$200$} & $0.1L/H$ & \multirow{5}{*}{$0.03$} & \multirow{5}{*}{139.30} & \multirow{5}{*}{0.089} & 17.61 & \multirow{5}{*}{3.52E-03} \\
         &  &  &  &  &  &  & $0.3L/H$ &  &  &  & 16.69 &  \\
         &  &  &  &  &  &  & $0.5L/H$ &  &  &  & 20.23 &  \\
         &  &  &  &  &  &  & $0.7L/H$ &  &  &  & 17.47 &  \\
         &  &  &  &  &  &  & $0.9L/H$ &  &  &  & 22.42 & 
        \end{tabular}
    \end{ruledtabular}
\end{table}

\bibliography{transport}